# Secondary Excitation of Spin-Waves: How Electromagnetic Cross-Talk Impacts on Magnonic Devices

Johannes Greil[1], Matthias Golibrzuch[1], Martina Kiechle[1], Ádám Papp[2],
Valentin Ahrens[1], György Csaba[2], and Markus Becherer[1]

[1] School of Computation, Information and Technology, Technical University of Munich, Germany
[2] Faculty of Information Technology and Bionics, Pázmány Péter Catholic University, Budapest, Hungary

This work examines the impact of electromagnetic cross-talk in magnonic devices when using inductive spin-wave (SW) transducers. We present detailed electrical SW spectroscopy measurements showing the signal contributions to be considered in magnonic device design. We further provide a rule of thumb estimation for the cross-talk that is responsible for the secondary SW excitation at the output transducer. Simulations and calibrated electrical characterizations underpin this method. Additionally, we visualize the secondary SW excitation via time-resolved MOKE imaging in the forward-volume configuration in a 100 nm Yttrium-Iron-Garnet (YIG) system. Our work is a step towards fast yet robust joint electromagentic-micromagnetic magnonic device design.

## I. Introduction

**M**AGNONICS has become a field of science in which the focus is no longer only on fundamental research but also on devices and their commercial usability. The promising properties of magnonic devices—tunability, ultra-low losses, GHz-regime applications—make them attractive for being miniaturized, integrated, and manufactured on a large scale. Despite the convenient inductive excitation and pick-up of spin-waves (SWs), the electromagnetic cross-talk between the SW transducers is a major hurdle when it comes to device design and benchmarking [1]. Additionally, the secondary excitation of SWs at the output transducer due to inductive coupling deteriorates the overall device performance.

An essential step between the functional design of a magnonic device and the layout of its electrical input/output (I/O) and matching network is the examination of the electromagnetic cross-talk that inevitably arises using inductive SW transducers. The parasitic cross-talk results in a significant modification of the transmitted SW signals in electrical measurements, most prominent in AESWS. While in magnon-transport evaluation the cross-talk is calibrated out via reference measurements at no or off-resonance DC magnetic bias field, the absolute magnitude of the direct coupling is crucially important for the practical use in real-time signal processing.

This work focuses on the origin of signal contributions of typical all-electrical SW spectroscopy (AESWS) [2]–[4] measurements in the forward-volume (FV) configuration [5]. These contributions can be explained by a straightforward cross-talk model based on the Hertzian dipole [6] for microstrip line (MSL) transducers and easy-to-access 2D simulations for coplanar waveguides (CPWs) using the magnetostatics solver FEMM [7]. Further, we provide time-resolved magneto-optical Kerr effect (trMOKE) images confirming the secondary SW excitation resulting from the electromagnetic cross-talk.

Corresponding author: J.Greil (email: johannes.greil@tum.de).

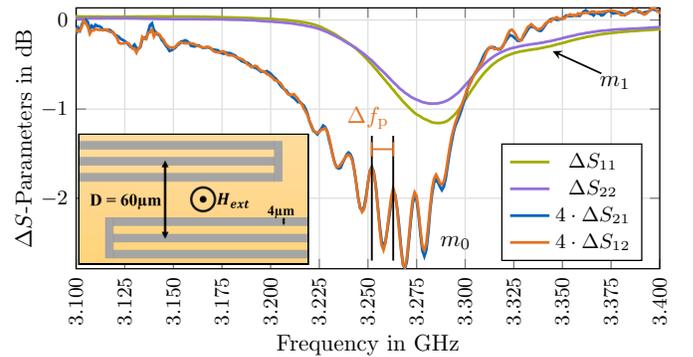

Fig. 1. $\Delta S$-Parameter for a CPW transducer pair with 60 μm center-to-center distance on a 100 nm YIG film in the forward-volume configuration (see top view sketch in the inset). The cross-talk of about 30 dB is subtracted. Both reflections show a dip at 3.285 GHz resulting from the power absorption in the YIG film. In transmission, the main signal contribution is still the power absorption indicated by the large dip in $\Delta S_{21}$ and $\Delta S_{12}$. The ripple on top of those signals is due to the varying spin-wave group velocity during a frequency sweep. Based on the calibrated VNA measurements it is possible to assess that the propagating SW signal has an amplitude of about 0.2 dB.

## II. Electrical Spin-Wave Spectroscopy Signal Contributions

Typical AESWS measurements in the FV configuration show three main signal contributions (in descending order of their amplitude): First, the electromagnetic cross-talk between the SW transducers due to the inductive coupling. Second, the absorption of RF power from the input transducer into the magnetic layer underneath. This absorption comes from the excitation of SWs that lowers the electromagnetic cross-talk measured via the transmission signals in the form of $S_{21}$ or $S_{12}$. Third, a signal contribution corresponding to SWs propagating from the input to the output transducer.

In Fig. 1 the latter two contributions are shown for the characterization of a 100 nm thin plain YIG film. The almost constant cross-talk of 40 dB in the evaluated frequency span is subtracted for better visualization ($\Delta S_{xx} = S_{xx,H_{ext}\neq 0} - S_{xx,H_{ext}=0}$). The inset shows the measurement configuration using a pair of shorted CPWs with equally broad signal (S) and ground (G) lines and gaps $w_\mathrm{S} = w_\mathrm{G} = w_\mathrm{gap} = 4\,\mu\mathrm{m}$, and a center-to-center distance of $D = 60\,\mu\mathrm{m}$. The transducers have a length of $1\,\mathrm{mm}$ and are tapered towards contact pads for bonding them to a chip carrier. The RF excitation frequency $f_\mathrm{RF}$ is swept from $3.1\,\mathrm{GHz}$ to $3.4\,\mathrm{GHz}$ with a step size of $625\,\mathrm{kHz}$ for an out-of-plane (OOP) DC magnetic bias field $\mu H_\mathrm{ext} = 260\,\mathrm{mT}$. The reflection signals $\Delta S_{11}$ and $\Delta S_{22}$ are used to determine the frequency of the most efficient SW excitation which is indicated by the dip at $3.285\,\mathrm{GHz}$ in both signals. Their different heights are caused by slightly varying bond connections to the carrier. A comparison to the transmission signal shows that the reflection signal peaks are located at about $2\,\mathrm{MHz}$ higher frequencies. The peak shift arises because in reflection measurements mainly the film characteristics directly underneath the CPW are probed, while in transmission the spin precession is enforced by additional but weaker field components up to several micron distances [6], [8]. Moreover, the first higher-order excitation mode $m_1$ of the CPW transducer—the first spatial harmonic of the cross-sectional field profile of the CPW geometry—generates another small dip in the reflection signals at $3.34\,\mathrm{GHz}$ [1], [8], [9].

The characterization of propagating SWs is obtained via the transmission parameters $\Delta S_{21}$ and $\Delta S_{12}$. Besides the subtracted electromagnetic cross-talk, the power absorption into the YIG film for exciting SWs is the strongest contribution and shows up as the large dip in $\Delta S_{21}$ and $\Delta S_{12}$ in Fig. 1. This implies that one can estimate the overall absorbed power along the transducer via calibrated VNA measurements. Especially in the FV geometry not more than half of the accepted power for SW excitation is transported towards the output transducer due to the isotropic wave propagation. For the example, from Fig. 1 we can determine that about $0.7\,\mathrm{dB}$, relative to the input power $P_\mathrm{in} = -10\,\mathrm{dBm}$, are absorbed into the film. The maximum amplitude of the ripple on top of the large dip is in turn $0.2\,\mathrm{dB}$. This ripple in $\Delta S_{21}$ and $\Delta S_{12}$ defines the third signal contribution which represents the propagating SW signal. The ripple stems from the changing SW wavelength while sweeping the excitation frequency for a fixed $H_\mathrm{ext}$ and constant transducer distance $D$ such that more SWs fit between the transducers. From the frequency spacing $f_\mathrm{p}$ between two adjacent peaks we can determine the group velocity via [3], [5], [9]

$$v_\mathrm{g} = \frac{\mathrm{d}\omega_\mathrm{SW}}{\mathrm{d}k} \approx \Delta f_\mathrm{p} D \quad (1)$$

to $v_\mathrm{g} \approx 660\,\mathrm{m/s}$. The first higher-order excitation of propagating spin-waves $m_1$ is also clearly visible in the transmission signal at $3.325\,\mathrm{GHz}$.

Summing up, calibrated AESWS measurements allow for a qualitative and quantitative distinction between the discussed signal contributions. Those device-specific characteristics set the basis for highly optimized magnonic devices, whereas the cross-talk reduction is the core challenge for a practically useful on-chip post processing of SW signals.

## III. Electromagnetic Cross-Talk Between Spin-Wave Transducers

The electromagnetic cross-talk is a severe signal contribution in the electrical evaluation of magnonic devices since it buries almost all SW-related signals. Thus, it is of great practical use to tackle the electromagnetic cross-talk in a straightforward and easy-to-use rule of thumb manner presented in the following.

The SW excitation using inductive transducers is described via the in-plane RF magnetic fields that force the spins to precess around the effective magnetic field $H_\mathrm{eff}$ [5]. This excitation is only applicable in very close proximity to the transducer lines because, as with conventional antennas, these field components correspond to the so-called reactive near field and decay fast compared to the far-field components [6]. For SW transducers with lengths in the range of several ten to hundred microns, the near-field character is preserved also for the electromagnetic cross-talk since for conventional, electrically small antennas the far-field distance $d_\mathrm{ff}$ is defined as [10]

$$d_\mathrm{ff} \approx 2\lambda_0. \quad (2)$$

In the near field, reactive power oscillates between the antenna structure and the surrounding space such that the Poynting vector is zero and no radiation takes place. In contrast, in the far field, a locally plane wave is formed because electric and magnetic field components are in phase [6]. From (2) we see, that $d_\mathrm{ff}$ is not dependent on the physical size of the transducer because it acts as a point-like source for the electromagnetic waves with comparably long wavelengths $\lambda_0$ in free-space. For applications up to $20\,\mathrm{GHz}$ the far field is thus reached first at $d_\mathrm{ff} \approx 30\,\mathrm{mm}$ which is much larger than a reasonable magnonic device dimension. The radiation characteristics of electrically small rod antennas can be modeled via the Hertzian Dipole (HD) if their cross-section is small compared to the transducer length [6]. The azimuth magnetic field component for $\vartheta = 90°$, i.e. in the $xy$-plane, of a HD aligned to the $z$-axis is written in polar coordinates as [6]

$$H_\varphi = \mathrm{i}\frac{kIl}{4\pi r}\left(1 + \frac{1}{\mathrm{i}kr}\right)e^{-\mathrm{i}kr}, \quad (3)$$

where $I$ is the impressed current, $l$ is the length of the dipole, $r$ is the radial distance to the origin, and $k$ is the (electromagnetic) wave number in azimuth direction. When flipping the HD over to be aligned with the $y$-axis, the azimuth field component describes the OOP field of an MSL in the $yx$-plane such that $H_\varphi \hat{=} H_z$. From (3) we see that for small values of $r$ the field amplitude decays in good approximation with $1/r^3$. In the far field, it decreases with $1/r$ as the second summand in (3) vanishes. Simulations in FEMM underpin that the HD is a good approximation for electrically short MSLs used as SW transducers. The widely used shorted CPW SW transducer [2], [8], [9], [11] can be constructed by a superposition of three such field components, written as $H_\mathrm{z,CPW} \approx -\frac{1}{2}H_\varphi(x+w) + H_\varphi(x) - \frac{1}{2}H_\varphi(x-w)$, where $w$ is the center distance between the signal line and the ground lines. The factors $-1/2$ stem from the fact that half the signal-line current returns in each of the ground lines. Thus, from the superposition, the near-field amplitude of a CPW resembles a decay proportional to $1/r^4$. While this rough approximation does not perfectly go in line with the simulated field profiles due to the neglected lateral dimension of the CPW, it offers a straightforward order-of-magnitude estimation of those.

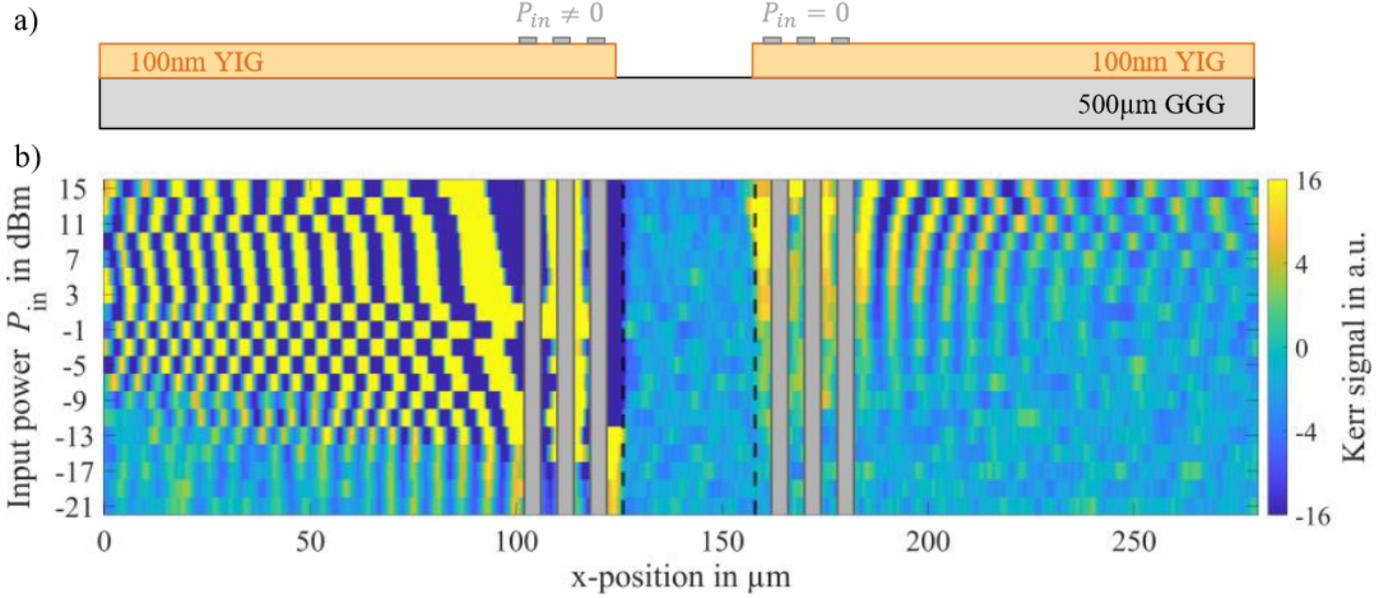

Fig. 2. a) Cross-sectional view of CPW transducers with $w_\text{S} = w_\text{gap} = w_\text{G} = 4\,\mu\text{m}$ and a center-to-center distance of $D = 60\,\mu\text{m}$ fabricated on 100 nm thin YIG. The SW channel between the transducers is removed over a width of $d = 35\,\mu\text{m}$. b) TrMOKE picture for a repeated scan of the same area, both on the input and output side, for input powers from $-21\,\text{dBm}$ to $15\,\text{dBm}$ at the supplied CPW (left). For the primarily excited SWs on the left the non-linear precession starts around $P_\text{in} = -15\,\text{dBm}$, while for the secondarily excited SWs $P_\text{in} \approx 5\,\text{dBm}$ is required. The non-linear excitation is visible by the power-dependent increase of the SW wavelength. The minimum input power to observe SWs is $-35\,\text{dBm}$ (not shown) and $-5\,\text{dBm}$ for the primary and secondary excitation, respectively.

As a consequence, the inductive coupling entails not only a strong cross-talk but also the secondary SW excitation at the output transducer. The shorted CPW forms two loops that pick up not merely the SW-induced signal but also the permeating magnetic out-of-plane fields from the input transducer. The resulting magnetic flux through the loop surface can be estimated via Faraday's law of induction

$$\Phi = \int_{A_\text{l}} B_\text{z} \cdot dA_\text{l}, \qquad (4)$$

where $A_\text{l} = 2w_\text{g}l$ is the area of the loop formed by the shorted CPW with length $l$ and gaps $w_\text{g}$. The induced voltage computes to $U_\text{ind} = -d\Phi/dt$. Further, the ohmic resistance of the electrically small CPW lines can be estimated via the general DC resistance model

$$R = \rho \frac{l}{A_\text{c}}, \qquad (5)$$

with the specific electrical resistivity $\rho$ and the cross-sectional area of a single line $A_\text{c}$. Hence, the induced current in the signal line is $I_\text{S,ind} = U_\text{ind}/R$, where by definition $I_\text{G,ind} = I_\text{S,ind}/2$ for a shorted CPW. Finally, approximating each of the CPW lines again as a thin conductor rod, the magnitude of the induced secondary IP magnetic field in the magnetic film underneath the output transducer can be estimated via Ampère's law

$$|H_\text{sec,IP}| \approx \frac{I_\text{ind}}{2\pi r}, \qquad (6)$$

where $r$ is the radial distance to the surface of the transducer line.

Let us discuss a specific example: for a pair of CPW transducers as shown in Fig. 3 with length $l = 1\,\text{mm}$, $w_\text{S} = w_\text{g} = w_\text{G} = 4\,\mu\text{m}$, a center-to-center distance $D = 60\,\mu\text{m}$, a metallization thickness of $300\,\text{nm}$, and an input power $P_\text{in} = -10\,\text{dBm}$ the following values can be estimated: For a remaining OOP magnetic field of about $6\,\mu\text{T}$ (FEMM simulation) at the output transducer forming two loops with an overall area of $A_\text{l} = 2w_\text{g}l = 0.008\,\text{mm}^2$, the induced voltage $U_\text{ind} \approx 30\,\text{mV}$. From (5) the induced current in 300 nm thick aluminum transducers is then $6\,\mu\text{A}$ which in turn induces a secondary IP magnetic field of $\mu_0 H_\text{sec,IP} \approx 20\,\mu\text{T}$ at half the thickness of the 100 nm YIG film. This secondarily induced IP field is sufficient to linearly excite SWs in the same manner as the supplied transducer does but at much lower power levels. In a perfect $50\,\Omega$ RF system this corresponds to an input power level of $P_\text{in} \approx -34\,\text{dBm}$ which is close to the minimum required input power in our trMOKE setup to resolve SW signals. In summary, the secondary excitation of SWs can be estimated in a rule-of-thumb manner via (4)-(6) showing reasonable results, also validated by the measurements in the following section. Thereby, the loop area $A_\text{l}$ of the CPW has a major influence on the strength of the secondary excitation.

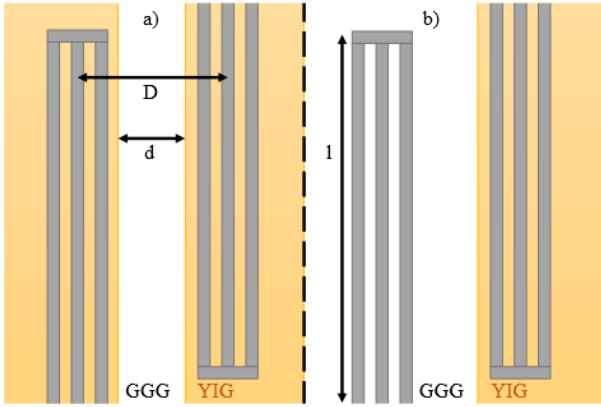

Fig. 3. Top view sketches of the sample settings discussed in Sec. IV. The distances $D = 60\,\mu\text{m}$ and $d = 35\,\mu\text{m}$ and the transducer length $l = 1\,\text{mm}$ are the same for both configurations. The trMOKE picture in Fig.2 corresponds to configuration a) for an excitation at the left transducer. The measurements shown in Fig. 4 and Fig. 5 correspond to configuration b) for an excitation at the right and left transducer, respectively.

## IV. SPATIAL VISUALIZATION OF SECONDARILY EXCITED SPIN-WAVES USING TIME-RESOLVED MOKE

### A. Both Transducers on YIG

To substantiate the presented method, we have considered sample settings that can be used to quantify the secondary SW excitation resulting from the electromagnetic cross-talk via trMOKE images. A first test configuration consists of two parallel shorted CPWs with length $l = 1\,\text{mm}$, $w_\text{S} = w_\text{gap} = w_\text{G} = 4\,\mu\text{m}$ and a center-to-center distance of $60\,\mu\text{m}$ fabricated on two areas of YIG separated by a $35\,\mu\text{m}$ wide gap as shown in Fig. 3 a. For $f_\text{RF} = 2.05\,\text{GHz}$ and $\mu H_\text{ext} \approx 214\,\text{mT}$, sweeping the input power $P_\text{in}$ from $-27\,\text{dBm}$ to $15\,\text{dBm}$ in steps of $2\,\text{dBm}$ results in the trMOKE image shown in Fig. 2. The left transducer is supplied by an RF source while the right one is not supplied and open at the tapered end. To avoid possible deviations arising from possible film inhomogeneities the same sample area is repeatedly scanned for the different input powers. Due to the large range of input powers, it is necessary to underscale the high-amplitude Kerr signals to resolve weaker signals. For the primary excitation, the minimum input power generating a resolvable Kerr signal is $-35\,\text{dBm}$, while non-linear excitation begins at around $P_\text{in} = -15\,\text{dBm}$. The transition to non-linear excitation is determined by the increase in wavelength compared to the scan at $P_\text{in} = -13\,\text{dBm}$. Looking at the secondarily excited SWs on the right, it is possible to detect wave fronts starting from $P_\text{in} \approx -5\,\text{dBm}$ at the supplied transducer. The non-linear secondary excitation starts at about $P_\text{in} = 5\,\text{dBm}$. In this configuration, the difference between the accepted power for primarily and secondarily excited SWs is about $10\,\text{dB}$ which corresponds to the overall cross-talk loss between the transducers. This loss contains the radiated but uncoupled field components of the supplied transducer that do not permeate the unsupplied CPW and the—in this sense unused—power that is absorbed at the left CPW for the primary excitation of SWs.

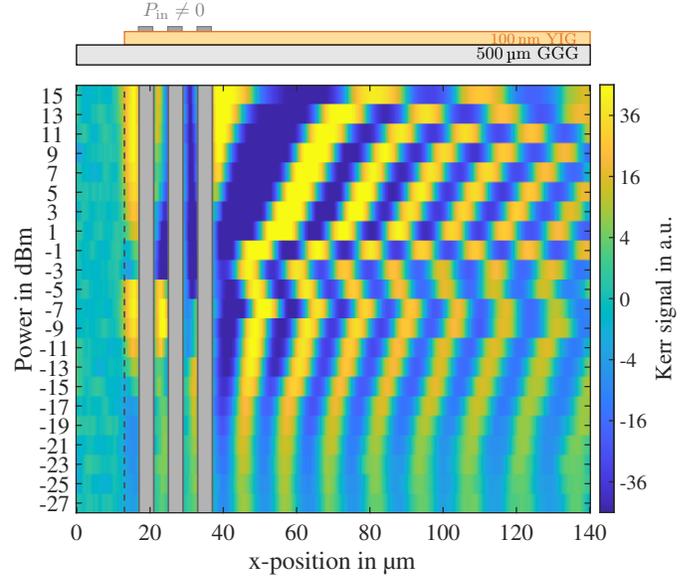

Fig. 4. TrMOKE picture of primarily excited SWs for an input power sweep from $-21\,\text{dBm}$ to $15\,\text{dBm}$ in steps of $2\,\text{dBm}$ at the right transducer in Fig. 3b. The lowest possible $P_\text{in}$ is $-35\,\text{dBm}$ (not shown), while the non-linear excitation starts at $P_\text{in} \approx -19\,\text{dBm}$ denoted by the increase in SW wavelength $\lambda_\text{SW}$. The intermediate bend of $\lambda_\text{SW}$ is attributed to a confinement effect between the sharp YIG edge and the CPW.

### B. One Transducer on GGG, One Transdcuer on YIG

A second configuration is shown in Fig. 3 b, where the left transducer is fabricated directly on the GGG substrate leaving only a $100\,\text{nm}$ YIG layer underneath the right transducer. The CPW dimensions are the same as for the previous measurement, the excitation frequency and external field are set to $f_\text{RF} = 2.05\,\text{GHz}$ and $\mu H_\text{ext} \approx 223\,\text{mT}$, respectively. The trMOKE picture of the primarily excited SWs is shown in Fig. 4 for input powers of $-27\,\text{dBm}$ to $15\,\text{dBm}$ in steps of $2\,\text{dBm}$. Similar to the former configuration a minimum input power of $-35\,\text{dBm}$ generates a resolvable Kerr signal, while the non-linear excitation now starts at around $P_\text{in} = -19\,\text{dBm}$. The intermediate bend of wavelengths for power levels between $-7\,\text{dBm}$ and $-1\,\text{dBm}$ is attributed to a confinement effect that occurs from the cavity-like structure formed by the sharp YIG edge and the CPW but is not yet investigated in detail.

The trMOKE picture of the secondarily excited SWs is shown in Fig. 5 for the same input power levels but this time applied to the transducer on the bare GGG substrate. Again, the same area is repeatedly scanned for all power levels to avoid influences of film inhomogeneities. The minimum required input power to detect SW signals is $-27\,\text{dBm}$, while the non-linear excitation begins at $P_\text{in} \approx 1\,\text{dBm}$. In this way we see that non-linearity is reached at $4\,\text{dB}$ less input power compared to the configuration in Sec. IV-A where both transducers are fabricated on YIG. This difference shows in turn how much power is absorbed into the YIG film for the primary SW excitation at the left transducer in Fig. 2. It reflects the fact that in the FV geometry a maximum of $50\,\%$ of the absorbed power can be transported by the spin waves and at least $50\,\%$ propagate in other directions. Thus, comparing the configurations in Fig. 3 a & b, we can measure the fraction of power that is absorbed for primary SW excitation and propagates away from the (removed) SW channel between the transducers.

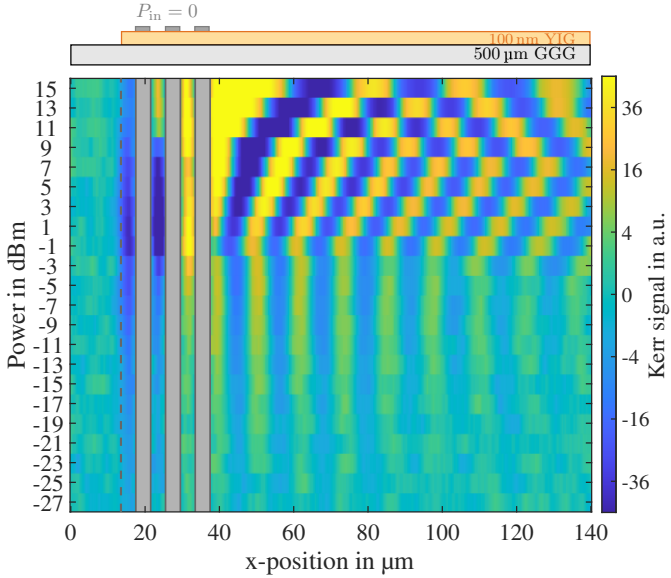

Fig. 5. TrMOKE picturce of secondarily excited SWs at the right CPW in Fig. 3b for an input power sweep at the left CPW from $-27\,\mathrm{dBm}$ to $15\,\mathrm{dBm}$ in steps of $2\,\mathrm{dBm}$. The minimum required input power to detect a SW signal is around $-23\,\mathrm{dBm}$, while the non-linear excitation starts at $P_\mathrm{in} \approx 1\,\mathrm{dBm}$.

## C. Distant Excitation of Spin-Waves

In a third configuration we omit the right transducer in Fig.3b such that only the left transducer is fabricated on the GGG substrate with distances of $20\,\mathrm{\mu m}$ to $60\,\mathrm{\mu m}$ to the YIG. In this experiment, it was not possible to excite SWs or near-FMR modes. It shows that there is not a distant FMR-like excitation [8] but the non-driven output transducer is necessary for the secondary SW excitation. At those relatively long distances between the transducer and the YIG layer the IP fields are too much confined underneath the supplied CPW to be strong enough to force the spins in the YIG to precession. Following the discussion in [12] e.g. an MSL in close proximity to the YIG edge could be a feasible transducer geometry to excite a relatively broad spectrum of SWs without the need to fabricate the transducer directly on YIG. However, to our knowledge the excitation efficiency of this method was not yet investigated experimentally. Even more, the fact that the secondary SWs are excited by the output transducer and not by a distant field component of the input CPW is indicated by the equally long SW wavelengths in the linear regime: From Fig. 2 we find that $\lambda_\mathrm{SW} \approx 7\,\mathrm{\mu m}$ for a linear excitation on both transducer sides. And from the comparison of Fig. 4 and Fig. 5 we find $\lambda_\mathrm{SW} \approx 10\,\mathrm{\mu m}$. The comparison shows that the wavelength for linear excitation depends on the cross-sectional geometry of the CPW, whereas the difference between the measured wavelengths in the two experiments results from the changed $H_\mathrm{ext}$ at a fixed $f_\mathrm{RF}$ [11].

## V. Conclusion

Our measurements show that using inductive SW transducers leads to two major challenges when it comes to high-speed signal transmission requiring lowest distortions: First, the electromagnetic cross-talk is the strongest contribution to the transmission signal and often buries the SW-related signals almost completely. The commonly used cross-talk subtraction in AESWS measurements is somewhat sufficient for investigating magnon-transport phenomena but is not applicable for real-time measurements in an I/O circuitry. And second, the electromagnetic cross-talk is such strong that we observe a secondary excitation of SWs at the output transducer due to the inductive coupling. These secondarily excited SWs interfere with the intentionally excited SWs and lower the magnonic signal-to-noise ratio as well as the overall detection sensitivity of the output transducer due to its own reaction on varying frequency or magnetic field. Thus, one of the main goals for paving the way for magnonic devices towards integration is to minimize the electromagnetic cross-talk between inductive SW transducers. This can be achieved either by shielding them from each other while maintaining the SW excitation efficiency as well as the electrical RF properties. Or by introducing asymmetric transducer structures, e.g. with a CPW as input and several loop antennas as outputs which are optimized for the magnonic device characteristics. A fully coupled electromagnetic-micromagnetic design approach is therefore essential for practically useful magnonic devices.

### Acknowledgment

The authors acknowledge funding from the European Union within HORIZON-CL4-2021-DIGITAL-EMERGING-01 (No. 101070536, MandMEMS), German Research Foundation (DFG No. 429656450), and the German Academic Exchange Service (DAAD, No. 57562081).

### References

[1] D. A. Connelly, G. Csaba et al., "Efficient electromagnetic transducers for spin-wave devices," *Scientific Reports*, vol. 11, no. 1, Sep. 2021.
[2] V. Vlaminck and M. Bailleul, "Current-induced spin-wave doppler shift," *Science*, vol. 322, no. 5900, pp. 410–413, Oct. 2008.
[3] S. Neusser, G. Duerr et al., "Anisotropic propagation and damping of spin waves in a nanopatterned antidot lattice," *Physical Review Letters*, vol. 105, no. 6, Aug. 2010.
[4] J. Chen, T. Yu et al., "Excitation of unidirectional exchange spin waves by a nanoscale magnetic grating," *Physical Review B*, vol. 100, no. 10, Sep. 2019.
[5] A. Prabhakar and D. D. Stancil, *Spin Waves: Theory and Applications*. Springer US, 2009.
[6] D. M. Pozar, *Microwave Engineering 4th Edition*. John Wiley & Sons, 2011.
[7] D. Meeker, "Finite element method magnetics." [Online]. Available: https://www.femm.info
[8] M. Sushruth, M. Grassi et al., "Electrical spectroscopy of forward volume spin waves in perpendicularly magnetized materials," *Physical Review Research*, vol. 2, no. 4, Nov. 2020.
[9] V. Vlaminck and M. Bailleul, "Spin-wave transduction at the submicrometer scale: Experiment and modeling," *Physical Review B*, vol. 81, no. 1, Jan. 2010.
[10] H. A. Wheeler, "The Radiansphere Around a Small Antenna," *Proceedings of the IRE*, vol. 47, Aug. 1959.
[11] J. Lucassen, C. F. Schippers et al., "Optimizing propagating spin wave spectroscopy," *Applied Physics Letters*, vol. 115, no. 1, p. 012403, Jul. 2019.
[12] Á. Papp, W. Porod et al., "Nanoscale spectrum analyzer based on spin-wave interference," *Scientific Reports*, vol. 7, no. 1, Aug. 2017.